\documentclass[ALICE,manyauthors]{cernphprep}

\usepackage[comma,square,numbers,sort&compress]{natbib}
\usepackage{hyperref}
\usepackage{lineno}
\usepackage{subfigure}
\usepackage[section]{placeins}
\usepackage {multicol}
\usepackage {multirow}
\usepackage {units}
\usepackage {array}
\usepackage {xcolor}
\usepackage{comment}
\usepackage{hyphenat}
\usepackage[T1]{fontenc}
\usepackage{orcidlink} 

\begin{document}

\begin{titlepage}

\PHyear{2026}       
\PHnumber{173}      
\PHdate{11 June}  
%

\title{Hyperon ($\Lambda$) polarization along the beam axis in Pb--Pb collisions at $\mathbf{\sqrt{s_{\mathrm{NN}}}=5.36}$~TeV}
\ShortTitle{$\Lambda$) polarization in Pb--Pb collisions}   

\Collaboration{ALICE Collaboration\thanks{See Appendix~\ref{app:collab} for the list of collaboration members}}
\ShortAuthor{ALICE Collaboration} 

\begin{abstract}
The measurement of hyperon ($\Lambda$ and $\overline{\Lambda}$) polarization along the beam axis in Pb--Pb collisions at $\sqrt{s_{\mathrm{NN}}} = 5.36$~TeV using the ALICE detector is presented. The polarization, arising from vorticity components induced by anisotropic flow, is studied relative to the second- and third-order event planes. The measured polarization exhibits clear azimuthal sine modulations, providing the first observation of polarization along the beam direction measured relative to the third-order event plane at the Large Hadron Collider. The values of the  polarization measured with respect to the second-order event plane relative to the second harmonic event plane are consistent with previous measurements at $\sqrt{s_{\mathrm{NN}}} = 5.02$~TeV and show improved statistical precision owing to the larger data sample. Comparisons with hydrodynamic calculations indicate that the measured polarization is sensitive to the bulk viscosity and the vortical structure of the quark--gluon plasma, offering new constraints on its transport properties. In particular, the polarization measured relative to the third-order event plane is expected to provide an additional and independent input to constrain both bulk and shear viscosities of the quark--gluon plasma.
\end{abstract}
 
\end{titlepage}

\setcounter{page}{2}


\section{Introduction}
\label{sec:intro}

In relativistic heavy-ion collisions, a strongly-interacting medium known as the quark--gluon plasma (QGP)~\cite{Adcox:2004mh, Adams:2005dq, ALICE:2022wpn} is formed. This medium exhibits collective behavior resembling that of a nearly perfect fluid and is characterized by nontrivial velocity and vorticity fields~\cite{Becattini:2007sr, Csernai:2014ywa}. These vortical structures can lead to the spin polarization of produced particles through spin-vorticity coupling~\cite{Liang:2004ph, Becattini:2015ska}. The observation of global polarization of $\Lambda$ hyperons at the Relativistic Heavy Ion Collider (RHIC)~\cite{STAR:2017ckg, STAR:2018gyt} and the Large Hadron Collider (LHC)~\cite{ALICE:2019onw} provided compelling evidence for the creation and evolution of the vortical fluid. The magnitude of the global polarization decreases with increasing collision energy, reflecting the reduced effect of baryon stopping at midrapidity~\cite{BRAHMS:2004adc}.

In addition to the global polarization arising from the system's orbital angular momentum, local vorticity fields may also emerge due to anisotropic flow~\cite{Betz:2007kg, Becattini:2017gcx, Sun:2018bjl, Voloshin:2017kqp}. The hydrodynamic-evolution model~\cite{Becattini:2017gcx, Palermo:2024tza}, transport model~\cite{Xia:2018tes, Fu:2020oxj}, and blast-wave model~\cite{Retiere:2003kf, STAR:2019erd, Voloshin:2017kqp} predict that the strong anisotropic flow developed in non-central collisions generates vorticity components along the beam axis, giving rise to a quadrupole and higher-order structure of the polarization in the transverse plane. The longitudinal polarization refers to the component of the polarization along the beam direction, which is adopted throughout this paper. The longitudinal polarization exhibits a modulation pattern similar to that of the collective flow, as both originate from the same underlying anisotropy of the medium.

Previous measurements by the STAR~\cite{STAR:2023eck, STAR:2019erd} and ALICE~\cite{ALICE:2021pzu} Collaborations in heavy-ion collisions revealed a non-zero longitudinal polarization with a positive sine-modulating component, opposite in sign to the prediction from thermal-vorticity-based models~\cite{Fu:2020oxj}. Recent theoretical developments incorporating shear-induced effects qualitatively reproduce the observed behavior and make predictions for centrality, which characterizes the degree of overlap between the two colliding nuclei, and transverse momentum ($p_{\mathrm{T}}$) dependences of the polarization signal~\cite{STAR:2019erd, Palermo:2024tza} for different parametrization of bulk and shear viscosities.

The measurement of longitudinal polarization measured relative to the high-order event planes is therefore of particular interest, as it can reveal how the underlying vorticity and transport properties of the QGP couple to different flow harmonics. Because the polarization is sensitive to both the initial-state geometry and the final-state collective dynamics~\cite{Niida:2024ntm}, it provides valuable constraints on the space--time evolution of the QGP. In particular, recent model calculations indicate that the longitudinal polarization at LHC energies is more sensitive to the bulk viscosity of the QGP than at RHIC energies~\cite{Palermo:2024tza}, thus offering a new handle on transport properties of the medium.

In this manuscript, the measurement of the longitudinal polarization of $\Lambda$ and $\overline{\Lambda}$ hyperons with respect to the second- and third-order event planes in Pb--Pb collisions at $\sqrt{s_{\mathrm{NN}}}=5.36$~TeV with the ALICE detector at the LHC is presented. Section~\ref{sec:setup} describes the experimental setup. In Sec.~\ref{sec:ana}, the analysis procedure is detailed. The systematic uncertainties in the measurement are described in Sec.~\ref{sec:syst}. Section~\ref{sec:results} presents results of the polarization measurement with model comparison. Conclusions are drawn in Sec.~\ref{sec:summary}.
\section{Experimental setup}
\label{sec:setup}
The data sample used in this analysis was collected with the ALICE detector during the 2023 heavy-ion data-taking period at the LHC. Details of the ALICE apparatus during the LHC Run~3 are described in Ref.~\cite{ALICE:2023udb}. The following detectors are used in the analysis: the Fast Interaction Trigger (FIT)~\cite{Antonioli:2013ppp}, which was newly installed for Run 3, as well as the Inner Tracking System (ITS)~\cite{ALICE:2013nwm} and the Time Projection Chamber (TPC)~\cite{ALICETPC:2020ann}, both upgraded for the LHC Run~3.

The FIT detector is composed of three subsystems, which are positioned on both sides of the interaction point in the forward-rapidity region. The FT0 consists of two arrays (FT0A and FT0C) of fast Cherenkov radiators, covering the full azimuthal angle within the pseudorapidity intervals $3.5 < \eta < 4.9$ and $-3.3 < \eta < -2.1$, respectively. The FV0 and FDD consist of scintillator arrays. The FV0 is located 3.5 m from the interaction point in the forward direction (corresponding to the positive-$z$ direction), while the FDD modules are installed on the forward and backward sides at distances of 17.0~m and 19.5~m, respectively. The FT0 detector is used to measure the minimum-bias interaction rate with the coincident signals of the FT0C and FT0A. In addition, the FT0C detector determines the centrality~\cite{ALICE:2025cjn} and event plane~\cite{ALICE:2013axi} in the final dataset. The collected and utilized number of events in the present analysis corresponds to approximately 5 billion, which corresponds to an integrated luminosity of $\mathcal{L}_{\mathrm{int}} \simeq $~3.2~nb$^{-1}$.

The ITS is composed of seven cylindrical layers of silicon detectors, which are arranged at radial distances between 2.3~cm to 39.3~cm from the beam pipe. The ITS is employed to determine the primary and secondary vertices and to reconstruct the charged-particle tracks close to the collision point. The primary-vertex position along the beam direction is required to be within 10~cm from the nominal interaction point. In addition, collisions are rejected if more than one reconstructed vertex is found in the same bunch crossing.

The TPC, which surrounds the ITS, is a large cylindrical detector equipped with a Gas Electron Multiplier (GEM)-based readout system, covering the longitudinal and radial dimension of $|z|<250$~cm and $85<r<247$~cm, respectively. The TPC reconstructs charged-particle tracks together with the ITS and identifies particle species based on the specific energy loss ($\mathrm{d}E/\mathrm{d}x$) of charged particles in the TPC gas.

\section{Data analysis}
\label{sec:ana}

The $\Lambda$ ($\overline{\Lambda}$) hyperons are reconstructed via the $\Lambda \rightarrow \mathrm{p}\pi^{-}$ ($\overline{\Lambda} \rightarrow \bar{\mathrm{p}}\pi^{+}$) decay channel. The decay products are identified using the $n\sigma_{\mathrm{TPC}}$ variable, defined as the difference between the measured ionization energy loss in the TPC and the expected value from the Bethe–Bloch parameterization for pions and protons, divided by the corresponding resolution. A selection of $|n\sigma_{\mathrm{TPC}}|<5$ is applied. In addition, decay products are required to cross at least 50 pad rows in the TPC to ensure good momentum resolution, and to satisfy $p_{\mathrm{T}}>$~0.2 (0.5)~GeV/$c$ and $|\eta|<$~0.8 for the pions (protons) to guarantee a uniform detector acceptance. Selected hyperons are further required to satisfy the topological selection described in Tab.~\ref{tab:v0sel} to enhance the signal significance with respect to the combinatorial background.

\begin{table}[h!]
\caption{Topological selection criteria for hyperon candidates}
\centering
\begin{tabular}{l|c}
\hline 
Selection variable & Requirement \\ \hline
Transverse decay radius & $>$~1.2~cm \\
Pion track DCA to primary vertex & $>$~0.05~cm \\ 
Proton track DCA to primary vertex & $>$~0.1~cm \\
Cosine of pointing angle & $>$~0.995 \\ 
Proper lifetime & $<$~30~cm \\
Rapidity & $|y| < 0.5$ \\
DCA between pion and proton & $<$~1$\sigma$ (resolution) \\ 
\hline 
\end{tabular}
\label{tab:v0sel}
\end{table}

The hyperon transverse decay radius is the distance between the primary vertex and the secondary vertex, while the distance of the closest approach (DCA) denotes the minimum distance between the decay product track and the primary vertex. The pointing angle is defined as the angle between the momentum of the reconstructed hyperon and the line connecting the primary and secondary vertices. The proper lifetime ($c\tau = Lm/p$) is calculated using momentum ($p$), invariant mass ($m$), and the distance between the primary vertex and the secondary vertex ($L$).

The hyperon polarization can be accessed through the parity-violating weak decay channel, in which the angular correlation between the $\Lambda$ and its decay product, the proton~\cite{Lee:1957qs}, can be expressed as
\begin{eqnarray}
\dfrac{\mathrm{d}N}{\mathrm{d}\Omega^{*}} = \dfrac{1}{4\pi} (1 + \alpha_{\mathrm{H}}P_{\mathrm{H}}\cos \theta^{*}),
\label{eq:angcor}
\end{eqnarray} 
where an asterisk denotes quantities defined in the hyperon rest frame, and $P_{\mathrm{H}}$ represents the hyperon polarization. $\theta^{*}$ is the angle between the polarization vector and the direction of the proton momentum. The hyperon decay parameter, which quantifies the strength of the parity-violating angular correlation between the hyperon spin and the momentum of its decay product, is represented as $\alpha_{\mathrm{H}}$ and is set to $\alpha_{\Lambda}= $~0.750~$\pm$~0.010 and $\alpha_{\overline{\Lambda}}=-0.758 \pm 0.012$~\cite{BESIII:2018cnd}.

Based on the angular correlations between the hyperon and its decay product, the polarization along the beam direction is accessible by measuring the proton angle relative to the beam direction in the hyperon rest frame for the $n$-th modulation, which can be expressed as
\begin{eqnarray}
P_{z, n} = \dfrac{\left< \cos \theta_{\mathrm{p}}^{*} \right>}{\alpha_{\mathrm{H}}\left< \cos^{2} \theta_{\mathrm{p}}^{*}\right>},
\end{eqnarray} 

where the $\theta_{\mathrm{p}}^{*}$ is the angle between the momentum of the proton and the beam direction in the hyperon rest frame. The term $\left< \cos^{2} \theta_{\mathrm{p}}^{*}\right>$ reflects the detector acceptance (Acc.) and is found to be $1/3$ for the ideal acceptance and decreases with decreasing $p_{\mathrm{T}}$~\cite{STAR:2023eck, ALICE:2021pzu}. Its dependence on rapidity and centrality is found to be small. All dependencies on $p_{\mathrm{T}}$, rapidity, and centrality are taken into account in the correction procedure.

Alternatively, the magnitude of the sine-modulating longitudinal polarization can be extracted by measuring the polarization multiplied by the sine function, which can be expressed as
\begin{eqnarray}
P_{z, sn} = \dfrac{\left< \cos \theta_{\mathrm{p}}^{*} \sin(n\varphi - n\Psi_{n}) \right>}{R_{n} \alpha_{H}\left< \cos^{2} \theta_{\mathrm{p}}^{*}\right> },
\end{eqnarray} 
where $\Psi_{n}$ is the reconstructed $n$-th event plane angle, and $R_{n}$ is the event-plane resolution~\cite{Poskanzer:1998yz}, using a three sub-event method to reduce the effects of longitudinal decorrelations. The event-plane resolution is evaluated to be 0.76 at 25\% centrality. 

The term $P_{z, sn}$ is extracted using an invariant-mass analysis, in which $\left< \cos \theta_{\mathrm{p}}^{*} \sin(n\varphi - n\Psi_{n}) \right>$ is evaluated as a function of the invariant mass for all hyperon candidates, following the procedure used in the previous publication~\cite{ALICE:2021pzu}. To separate the signal from the background, a simultaneous fit to the invariant mass and polarization distributions is performed. The polarization distribution is fitted using the equation:
\begin{eqnarray}
P_{\mathrm{tot}}(M_{\mathrm{inv}}) = f_{\mathrm{sig}}P_{\mathrm{sig}}(M_{\mathrm{inv}}) + (1-f_{\mathrm{sig}})P_{\mathrm{bkg}}(M_{\mathrm{inv}}),
\end{eqnarray} 
where $f_{\mathrm{sig}}$ denotes the signal fraction, and $P_{\mathrm{sig}}$ and $P_{\mathrm{bkg}}$ represent the polarization signal and background, respectively. In this procedure, $f_{\mathrm{sig}}$ is used as a shared parameter between the invariant-mass and polarization fits.

\begin{figure}[!hbt]
	\centering
	\subfigure{ \includegraphics[width=0.6 \textwidth]{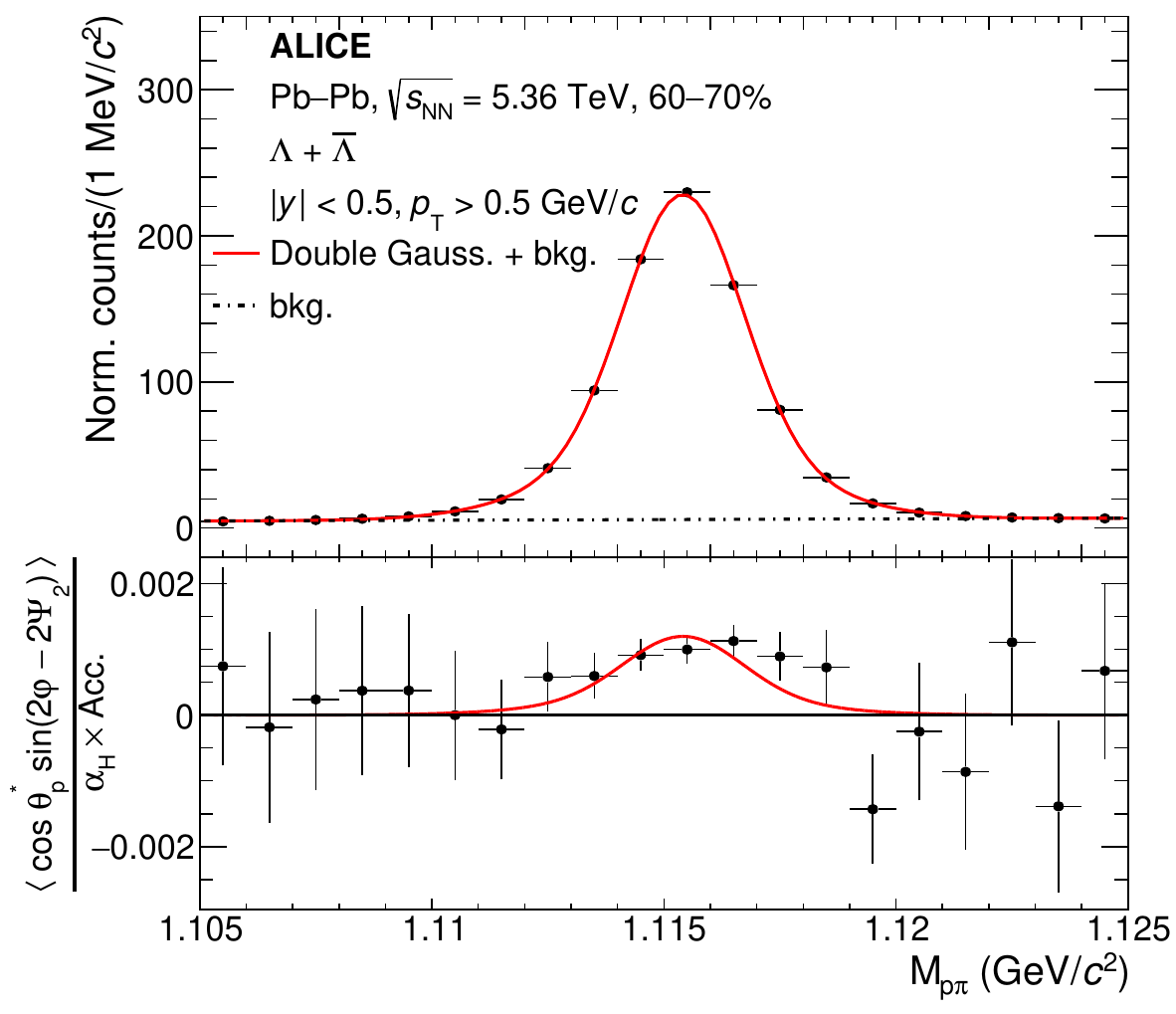} }
	\caption{Invariant-mass (top panel) and polarization (bottom panel) distributions of $\Lambda + \overline{\Lambda}$ in the 60--70\% Pb--Pb collisions at $\sqrt{s_{\mathrm{NN}}} = 5.36$~TeV.}
	\label{fig:sigext}
\end{figure}

Figure~\ref{fig:sigext} shows the invariant-mass and the polarization distributions on the upper and lower panel, respectively, in the 60--70\% central Pb--Pb collisions at $\sqrt{s_{\mathrm{NN}}} = 5.36$~TeV. The signal and background distributions are modeled with a double Gaussian and a third-order polynomial function, respectively. The polarization is corrected for the detector acceptance and the decay parameter $ \alpha_{H}$. The background polarization $P_{\mathrm{bkg}}$ is consistent with zero and shows no significant dependence on the order of the polynomial function.

\section{Systematic uncertainties}
\label{sec:syst}

The systematic uncertainties of longitudinal hyperon polarization are estimated by varying the analysis selection criteria, event-plane determination, and fitting procedure, as well as feed-down effects for hyperons. Each systematic variation is performed solely while other selections remain unchanged. A Barlow test~\cite{Barlow:2002yb} is performed for each systematic variation, requiring that the standard deviation of the Barlow factor distribution does not exceed 1.3. Most of the relative uncertainties are large only in central collisions, where the polarization is close to zero.

The same analysis is conducted with a different primary-vertex selection of $|z_\mathrm{vtx}|<$~8~cm with respect to the default selection of $|z_\mathrm{vtx}|<$~10~cm to evaluate the systematic uncertainty from the primary-vertex selection. The systematic deviation from the variation of the selection is found to be negligible compared to the statistical fluctuations, and therefore no uncertainty is assigned. The systematic uncertainty from the event-plane detector is evaluated by replacing the FT0C with TPC tracks at forward and backward rapidity. The systematic uncertainty from the event-plane detector is found to be negligible.

The systematic uncertainty on the hyperon selection is evaluated by varying the topological requirements for kinematic variables of reconstructed hyperons. For each selection, the variations are defined by applying both stricter and looser criteria with respect to the default values listed in Tab.~\ref{tab:v0sel}. The resulting systematic deviation from these variations yields the systematic uncertainties of 1--8\% and 2--10\% for the second- and third-order polarization, respectively, from peripheral to central collisions. The systematic uncertainty on the selection for decay products from hyperons is calculated by varying requirements for the minimum number of crossed TPC pad rows and the measured d$E$/d$x$ deviation. Systematic uncertainties are found to be 1--12\% and 1--20\% for the second- and third-order polarization, respectively, from peripheral to central collisions.

Systematic uncertainties associated with the signal extraction are estimated by changing the background modeling from second- to third-order polynomial function. The systematic deviation is found to be negligible from the variation of the background function. The systematic uncertainties associated with the feed-down contribution to $\Lambda$ polarization are conservatively estimated to be 10\%, based on the studies~\cite{Xia:2019fjf, Becattini:2019ntv} for second- and third-order polarizations.


\section {Results}
\label{sec:results}

\begin{figure}[!hbt]
	\centering
	\subfigure{ \includegraphics[width=0.48 \textwidth]{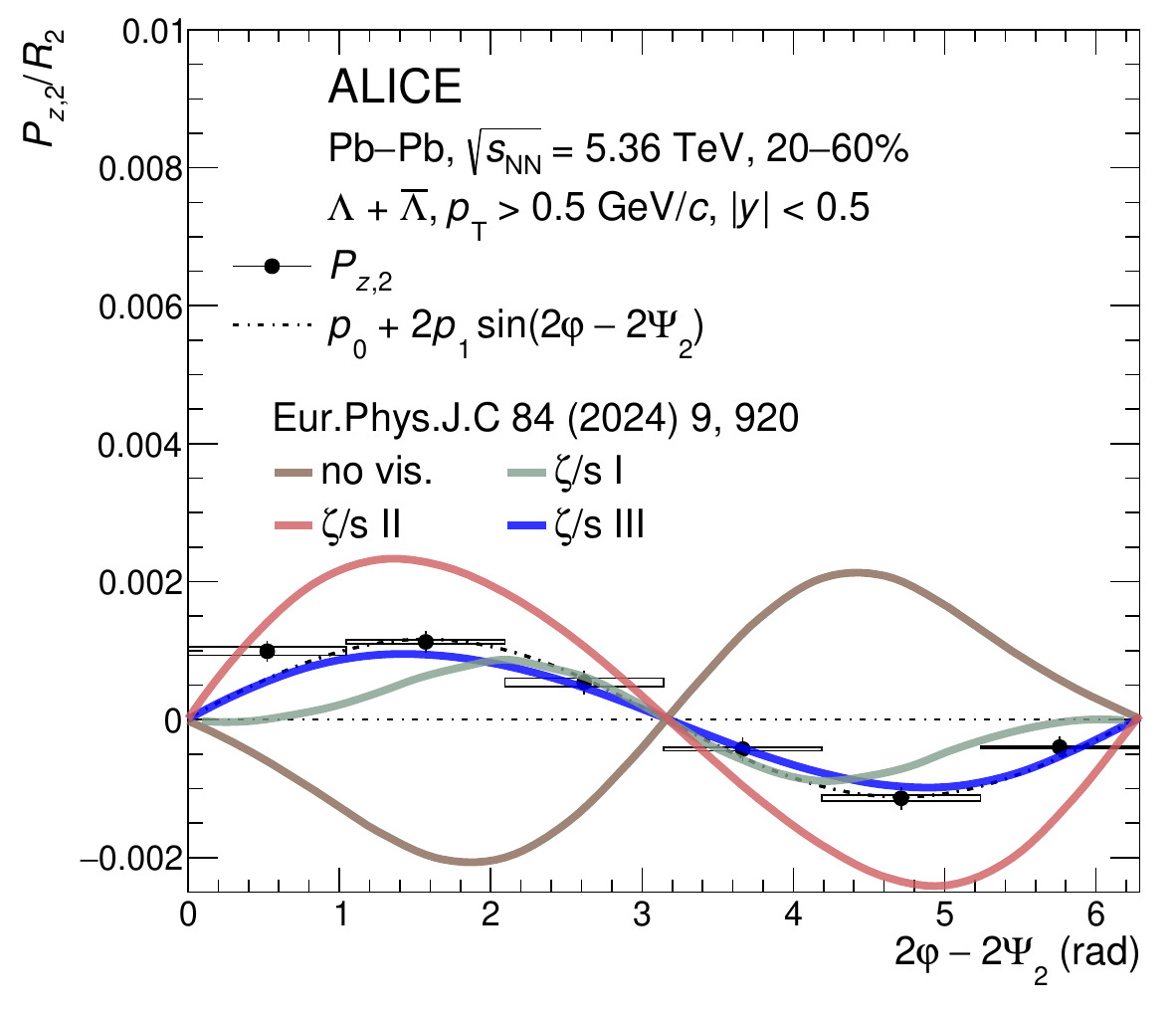} }
	\subfigure{ \includegraphics[width=0.48 \textwidth]{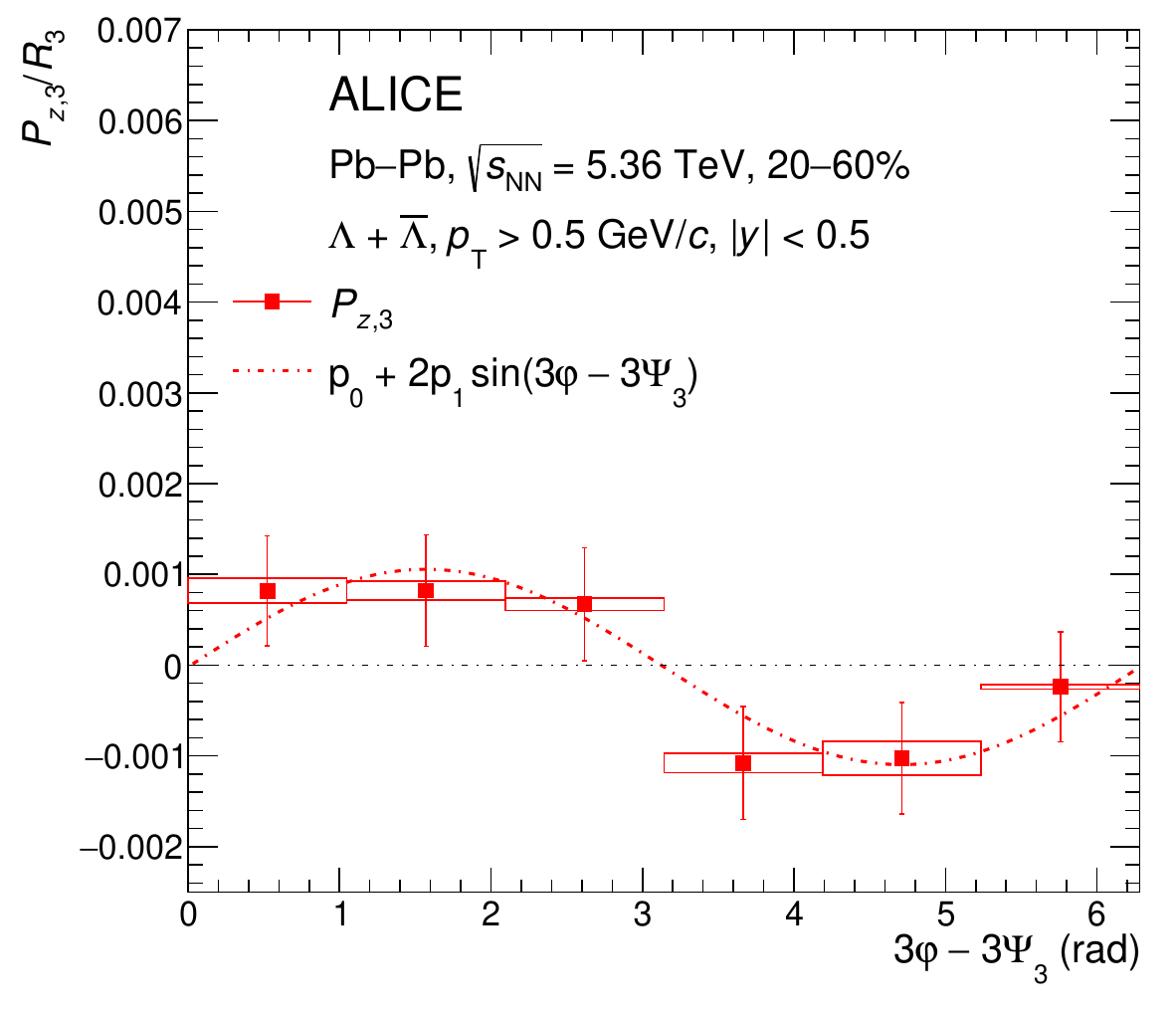} }
	\caption{Longitudinal hyperon polarization as a function of relative azimuthal angle in 20--60\% Pb--Pb collisions at $\sqrt{s_{\mathrm{NN}}} = 5.36$~TeV for the second-order (left) and third-order (right) modulations. Statistical and systematic uncertainties are represented as bars and boxes, respectively. The distributions are fitted with second- and third-order sine functions for the left and right panels, respectively. The solid lines from the left panel represent model calculations of the longitudinal polarization for different viscosity implementations from Ref.~\cite{Palermo:2024tza}.}
	\label{fig:poldphi}
\end{figure}

Figure~\ref{fig:poldphi} shows the longitudinal polarization as a function of the azimuthal angle relative to the second- and third-order event planes in Pb--Pb collisions at $\sqrt{s_{\mathrm{NN}}} = 5.36$~TeV with the correction for the event-plane resolution. The polarization is corrected for the event-plane resolution of the FT0C and is fitted with sine modulations. The value of parameter of $p_{0}$ is found to be zero, and the parameter of $p_{1}$ corresponds to the magnitude of the longitudinal polarization. The magnitudes of the second- and third-order polarizations are comparable, in contrast to the typical ordering observed for the flow coefficients. 

The model calculations are based on a hybrid framework combining initial conditions, (3+1)D viscous hydrodynamic evolution, particlization, and hadronic transport~\cite{Palermo:2024tza}. Within this framework, the polarization is evaluated at the hadronization hypersurface assuming local thermodynamic equilibrium. The resulting spin polarization arises from multiple contributions related to fluid gradients, including vorticity, shear, acceleration, and expansion effects. Therefore, the observed polarization reflects the interplay and partial cancellation of these different contributions, rather than a pure manifestation of vorticity alone. In this context, the relative importance of these contributions is strongly influenced by the transport properties of the medium. The experimental polarization can be explained with the QGP bulk viscosity $(\zeta/\mathrm{s})$ implementation. With the absence of the QGP bulk viscosity, the calculated longitudinal polarization is opposite to the experimental data. Different bulk-viscosity options are implemented via phenomenological temperature-dependent parametrizations that feature a peak around the critical temperature, reflecting the expected enhancement of bulk viscosity near the QCD phase transition~\cite{Schenke:2020mbo, Ryu:2017qzn, Schenke:2019ruo}. 

The bulk viscosity plays a particularly important role in the longitudinal polarization at LHC energies. While its effect is relatively modest at lower collision energies such as those at RHIC, it becomes significant at LHC energies, where the inclusion of bulk viscosity leads to a sign change of the longitudinal polarization. This behavior arises because bulk viscosity modifies the space–time evolution of the system and, consequently, the velocity gradients entering the polarization mechanism. 
Although it does not directly generate polarization, it changes the relative balance among different contributions, leading to a reorganization of their interplay. As a result, the inclusion of bulk viscosity can qualitatively alter the polarization signal at LHC energies, bringing the calculations into agreement with the observed trend. This sensitivity indicates that longitudinal polarization provides constraints not only on the vortical structure of the medium, but also on its transport properties.

\begin{figure}[!hbt]
	\centering
	\subfigure{ \includegraphics[width=0.6 \textwidth]{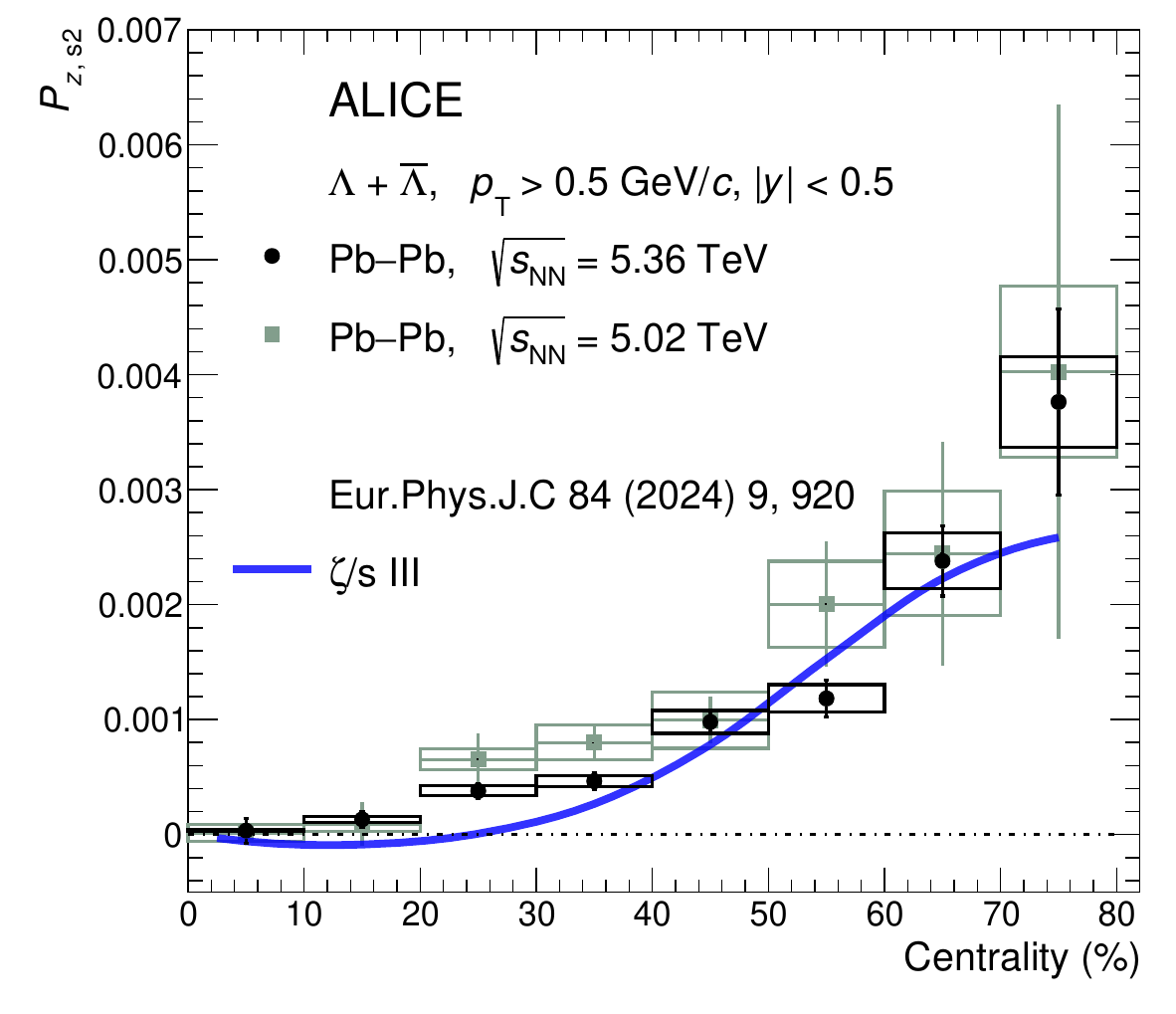} }
	\caption{Second-order longitudinal hyperon polarization as a function of centrality in Pb--Pb collisions at $\sqrt{s_{\mathrm{NN}}} = 5.36$~TeV and 5.02~TeV. Statistical and systematic uncertainties are represented as bars and boxes, respectively. The solid line represents model calculations for the longitudinal polarization with the viscosity implementation (Option III) from Ref.~\cite{Palermo:2024tza}.}
	\label{fig:pol2m}
\end{figure}

Figure~\ref{fig:pol2m} shows the second harmonic amplitude of the longitudinal hyperon polarization as a function of centrality in Pb--Pb collisions at $\sqrt{s_{\mathrm{NN}}} = 5.36$~TeV. The polarization increases monotonically with increasing centrality percentile. The measured polarization is compared with the polarization in Pb--Pb collisions at $\sqrt{s_{\mathrm{NN}}} = 5.02$~TeV. The two measurements agree within at most two standard deviations while the statistical uncertainties are improved in Pb--Pb collisions at $\sqrt{s_{\mathrm{NN}}} = 5.36$~TeV due to a larger data sample. The increasing trend can be understood as a consequence of larger initial geometry anisotropy and velocity gradients in more peripheral collisions. The centrality dependence is qualitatively reproduced by the model calculation with the implementation of the QGP viscosity~\cite{Palermo:2024tza}.

\begin{figure}[!hbt]
	\centering
	\subfigure{ \includegraphics[width=0.6 \textwidth]{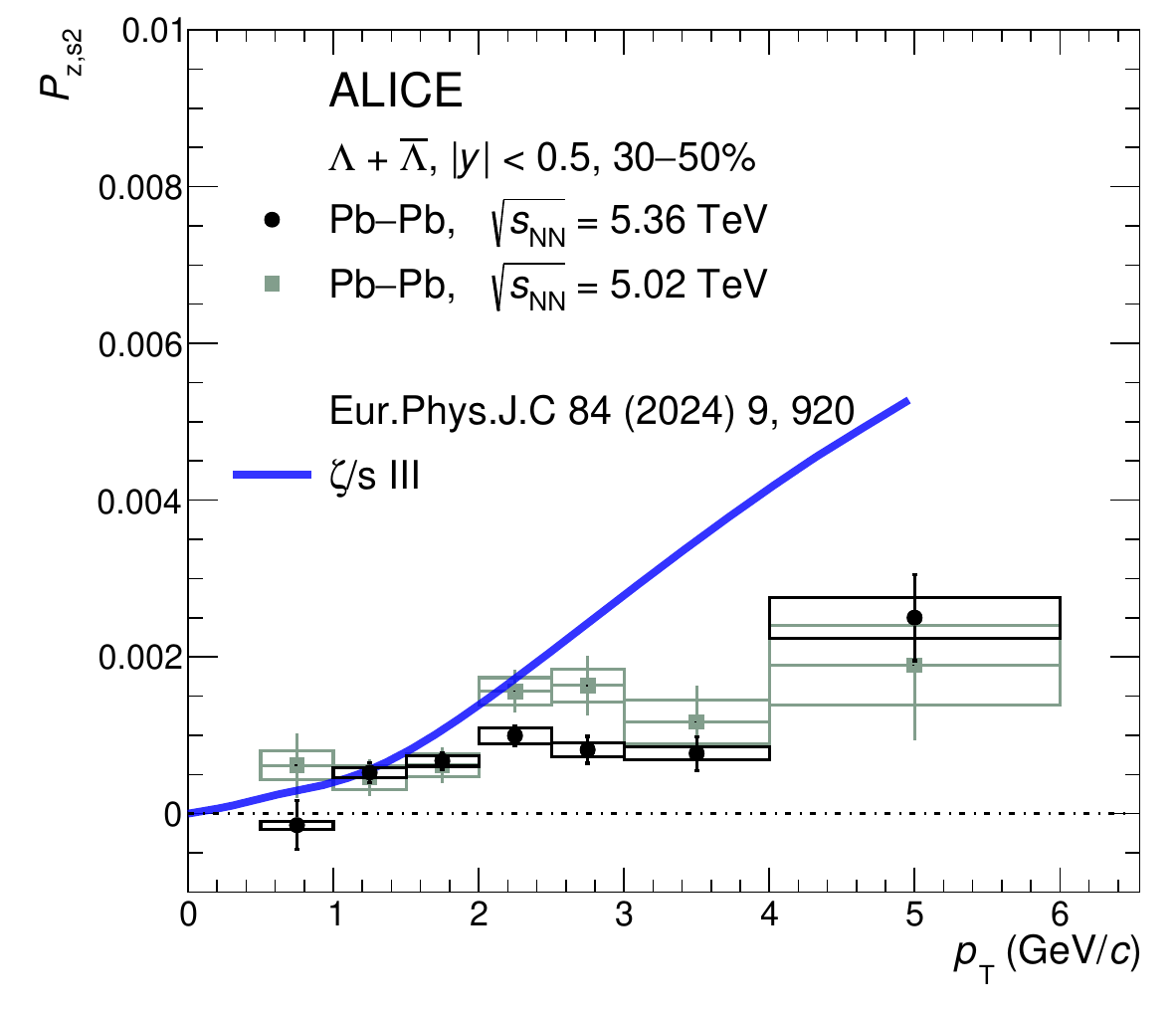} }
	\caption{Second-order longitudinal hyperon polarization as a function of $p_{\mathrm{T}}$ in the 30--50\% Pb--Pb collisions at $\sqrt{s_{\mathrm{NN}}} = 5.36$~TeV and 5.02~TeV. Statistical and systematic uncertainties are represented as bars and boxes, respectively. The solid line represents model calculations for the longitudinal polarization with the viscosity implementations (Option III) from Ref.~\cite{Palermo:2024tza}.}
	\label{fig:pol2p}
\end{figure}

Figure~\ref{fig:pol2p} shows the second harmonic amplitude of the longitudinal hyperon polarization as a function of $p_{\mathrm{T}}$ in the 30--50\% Pb--Pb collisions at $\sqrt{s_{\mathrm{NN}}} = 5.36$~TeV. The polarization is weakly increasing with increasing $p_{\mathrm{T}}$. The measured polarization is also compared with results obtained in Pb--Pb collisions at $\sqrt{s_{\mathrm{NN}}} = 5.02$~TeV. The statistical uncertainties are improved in Pb–Pb
collisions at $\sqrt{s_{\mathrm{NN}}} = 5.36$~TeV, benefiting from the larger data sample. The improved statistical precision of the Run 3 data provides physics input beyond a consistency check with the earlier Run 2 measurement. In particular, the present data indicate a weak transverse-momentum dependence of the second-order longitudinal polarization, with a tendency toward saturation at intermediate $p_{\rm T}$. The $p_{\mathrm{T}}$ dependence is not fully reproduced by current viscous hydrodynamic calculations, which tend to overestimate the increase of the signal at high $p_{\rm T}$. The measurement therefore begins to discriminate between different dynamical descriptions of the spin-polarization mechanism.

\begin{figure}[!hbt]
	\centering
	\subfigure{ \includegraphics[width=0.48 \textwidth]{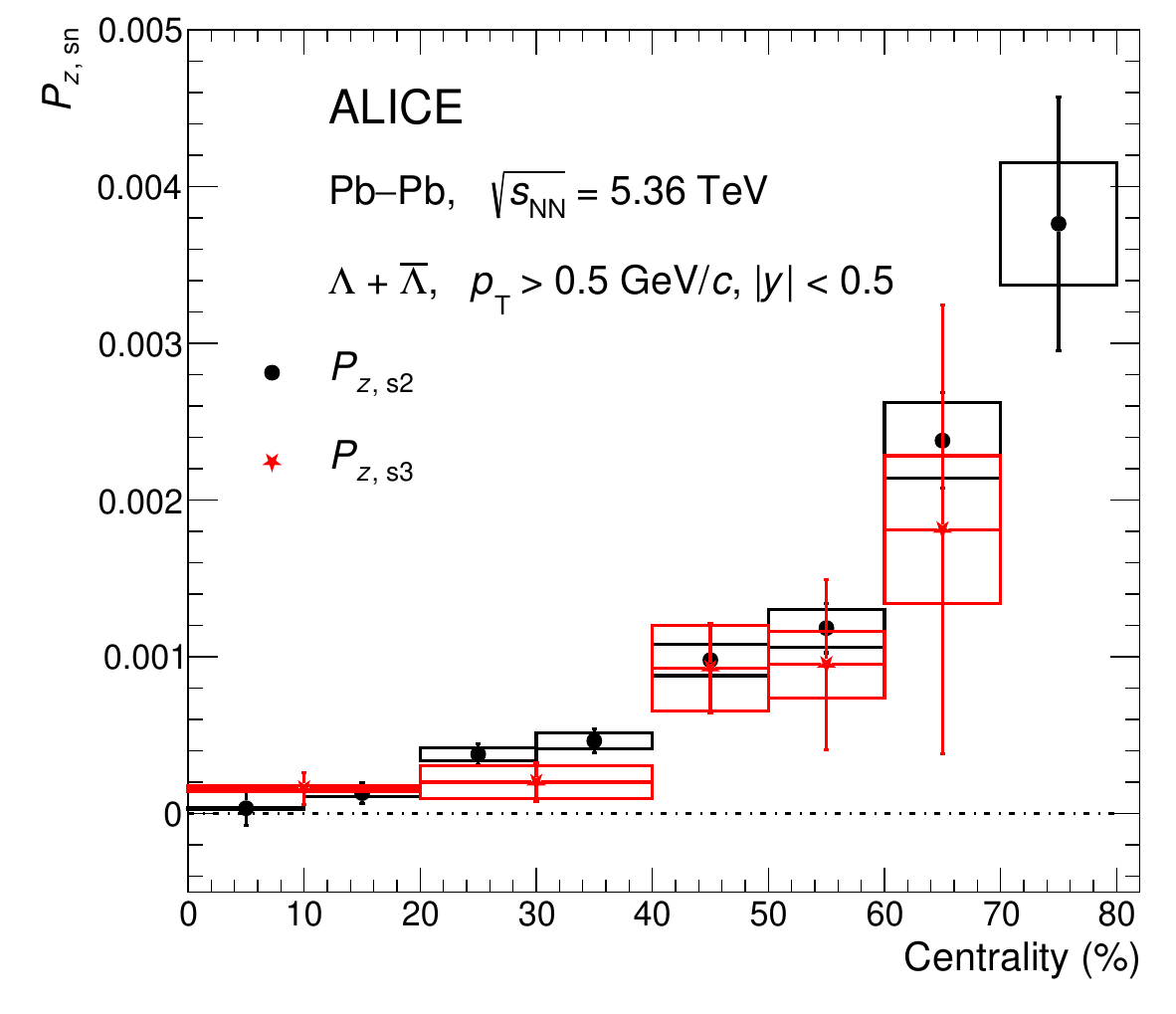} }
    \subfigure{ \includegraphics[width=0.48 \textwidth]{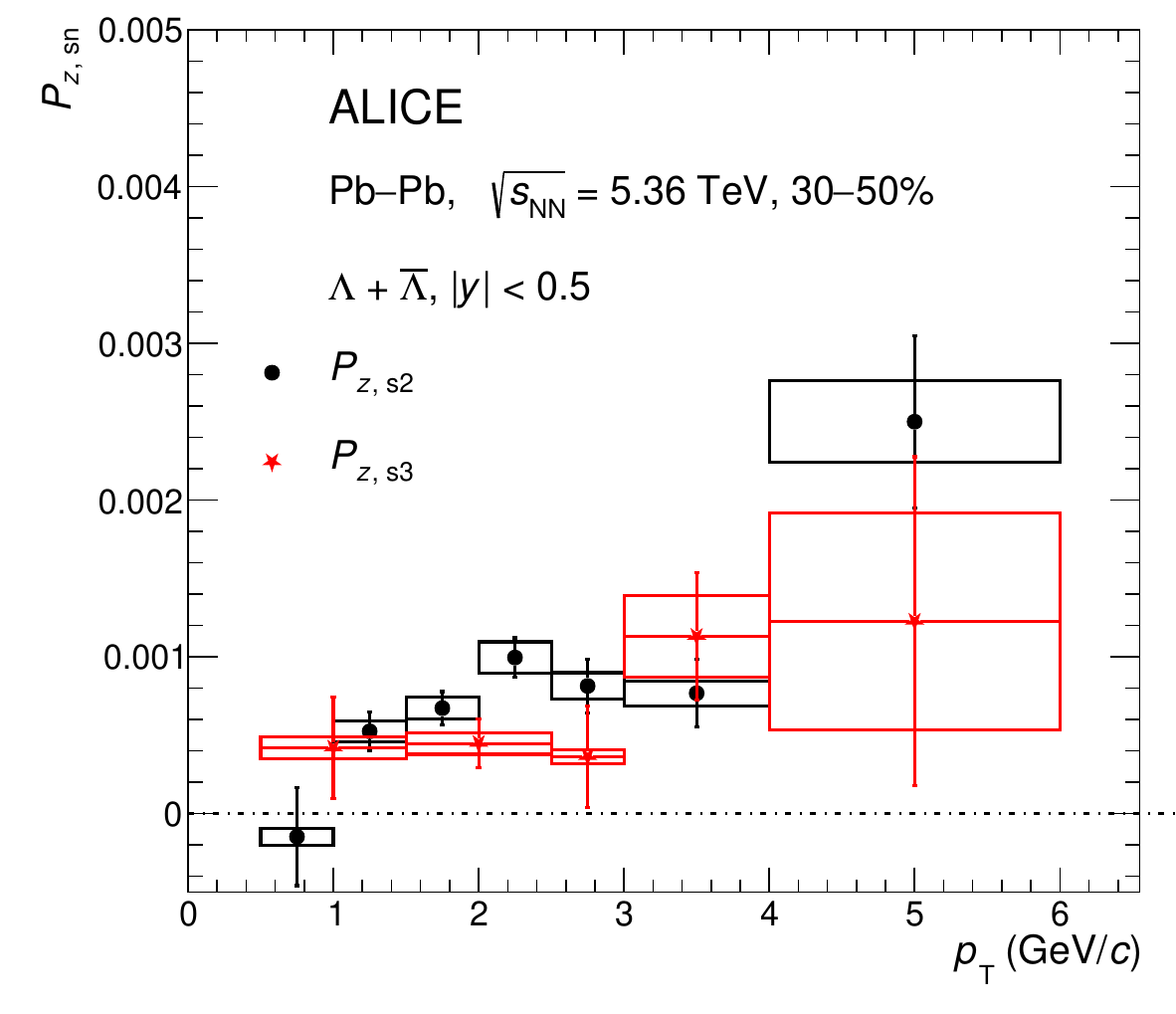} }
	\caption{Second- and third-order longitudinal hyperon polarization as a function of centrality (left) and $p_{\mathrm{T}}$ (right) in Pb--Pb collisions at $\sqrt{s_{\mathrm{NN}}} = 5.36$~TeV. Statistical and systematic uncertainties are represented as bars and boxes, respectively.}
	\label{fig:pol3}
\end{figure}
In addition, the first LHC measurement of the third-order longitudinal polarization provides qualitatively new information on fluctuation-driven vortical structures associated with higher-order anisotropic flow.
Figure~\ref{fig:pol3} shows the longitudinal hyperon polarization relative to the second- and third-order event planes as a function of centrality and $p_{\mathrm{T}}$ in Pb--Pb collisions at $\sqrt{s_{\mathrm{NN}}} = 5.36$~TeV in the left and right panels, respectively. The fully corrected polarizations are comparable in magnitude between the second- and third-order modulations, whereas the corresponding flow coefficients are systematically higher for the second order from semi-central collisions onward~\cite{ALICE:2016ccg, ALICE:2014wao}. The comparison of data with model calculations for the third-order polarization is expected to provide further constraints on the QGP viscosity.

In particular, the first LHC measurement of the third-order longitudinal polarization provides qualitatively new information on fluctuation-driven vortical structures associated with higher-order anisotropic flow. While the second-order component is mainly driven by the average elliptic geometry of the system, the third-order polarization is sensitive to event-by-event initial-state fluctuations associated with triangular flow. The combined measurement of second- and third-order polarization therefore offers a more differential constraint on the underlying mechanism, probing both the average geometry and fluctuation-driven vortical structures of the medium.


\section{Conclusions}
\label{sec:summary}

High-precision measurements of longitudinal hyperon ($\Lambda$ and $\overline{\Lambda}$) polarization in Pb--Pb collisions at $\sqrt{s_{\mathrm{NN}}} = 5.36$~TeV are presented. The longitudinal hyperon polarization relative to the second-order event plane is observed to increase toward peripheral collisions and to show a mild dependence on transverse momentum, in qualitative agreement with earlier measurements at  $\sqrt{s_{\mathrm{NN}}} = 5.02$~TeV. Benefiting from the large Run~3 data sample, the present results feature substantially reduced statistical uncertainties, allowing for more detailed and stringent comparisons with theoretical models. Comparisons with recent viscous hydrodynamic calculations indicate that the measured polarization is sensitive to the transport properties of the QGP. In particular, the inclusion of the bulk viscosity is found to be important to reproduce the sign of the observed longitudinal hyperon polarization relative to the second-order event plane, underscoring the role of dissipative effects in the development of the local spin polarization in the medium.

In addition, the measurement of the longitudinal hyperon polarization relative to the third-order event plane is performed for the first time at the LHC. These observables provide an independent and complementary handle on the vortical structure of the system, which is largely influenced by event-by-event initial-state fluctuations and higher-order anisotropic flow. The third-order component therefore extends the sensitivity of longitudinal polarization measurements beyond the dominant elliptic geometry.

Recent theoretical studies motivate the combined investigation of multiple harmonic components of longitudinal polarization, as such observables can receive contributions from different mechanisms, including thermal vorticity and shear-induced effects, depending on the viscous properties of the medium and the initial-state geometry. In this context, higher-order polarization measurements offer additional information that is complementary to that obtained from traditional flow observables. Therefore, the present measurement provides an important opportunity to constrain the vortical response and transport properties of the QGP through the study of higher-order polarization structures.

\newenvironment{acknowledgement}{\relax}{\relax}
\begin{acknowledgement}
\section*{Acknowledgements}
\noindent 

The ALICE Collaboration would like to thank all its engineers and technicians for their invaluable contributions to the construction of the experiment and the CERN accelerator teams for the outstanding performance of the LHC complex.
The ALICE Collaboration gratefully acknowledges the resources and support provided by all Grid centres and the Worldwide LHC Computing Grid (WLCG) collaboration.
The ALICE Collaboration acknowledges the following funding agencies for their support in building and running the ALICE detector:
A. I. Alikhanyan National Science Laboratory (Yerevan Physics Institute) Foundation (ANSL), State Committee of Science and World Federation of Scientists (WFS), Armenia;
Austrian Academy of Sciences, Austrian Science Fund (FWF): [M 2467-N36] and Nationalstiftung f\"{u}r Forschung, Technologie und Entwicklung, Austria;
Ministry of Communications and High Technologies, National Nuclear Research Center, Azerbaijan;
Rede Nacional de Física de Altas Energias (Renafae), Financiadora de Estudos e Projetos (Finep), Funda\c{c}\~{a}o de Amparo \`{a} Pesquisa do Estado de S\~{a}o Paulo (FAPESP) and The Sao Paulo Research Foundation  (FAPESP), Brazil;
Bulgarian Ministry of Education and Science, within the National Roadmap for Research Infrastructures 2020-2027 (object CERN), Bulgaria;
Ministry of Education of China (MOEC) , Ministry of Science \& Technology of China (MSTC) and National Natural Science Foundation of China (NSFC), China;
Ministry of Science and Education and Croatian Science Foundation, Croatia;
Centro de Aplicaciones Tecnol\'{o}gicas y Desarrollo Nuclear (CEADEN), Cubaenerg\'{\i}a, Cuba;
Ministry of Education, Youth and Sports of the Czech Republic, Czech Republic;
The Danish Council for Independent Research | Natural Sciences, the VILLUM FONDEN and Danish National Research Foundation (DNRF), Denmark;
Helsinki Institute of Physics (HIP), Finland;
Commissariat \`{a} l'Energie Atomique (CEA) and Institut National de Physique Nucl\'{e}aire et de Physique des Particules (IN2P3) and Centre National de la Recherche Scientifique (CNRS), France;
Bundesministerium f\"{u}r Forschung, Technologie und Raumfahrt (BMFTR) and GSI Helmholtzzentrum f\"{u}r Schwerionenforschung GmbH, Germany;
National Research, Development and Innovation Office, Hungary;
Department of Atomic Energy Government of India (DAE), Department of Science and Technology, Government of India (DST), University Grants Commission, Government of India (UGC) and Council of Scientific and Industrial Research (CSIR), India;
National Research and Innovation Agency - BRIN, Indonesia;
Istituto Nazionale di Fisica Nucleare (INFN), Italy;
Japanese Ministry of Education, Culture, Sports, Science and Technology (MEXT) and Japan Society for the Promotion of Science (JSPS) KAKENHI, Japan;
Consejo Nacional de Ciencia (CONACYT) y Tecnolog\'{i}a, through Fondo de Cooperaci\'{o}n Internacional en Ciencia y Tecnolog\'{i}a (FONCICYT) and Direcci\'{o}n General de Asuntos del Personal Academico (DGAPA), Mexico;
Nederlandse Organisatie voor Wetenschappelijk Onderzoek (NWO), Netherlands;
The Research Council of Norway, Norway;
Pontificia Universidad Cat\'{o}lica del Per\'{u}, Peru;
Ministry of Science and Higher Education, National Science Centre and WUT ID-UB, Poland;
Korea Institute of Science and Technology Information and National Research Foundation of Korea (NRF), Republic of Korea;
Ministry of Education and Scientific Research, Institute of Atomic Physics, Ministry of Research and Innovation and Institute of Atomic Physics and Universitatea Nationala de Stiinta si Tehnologie Politehnica Bucuresti, Romania;
Ministerstvo skolstva, vyskumu, vyvoja a mladeze SR, Slovakia;
National Research Foundation of South Africa, South Africa;
Swedish Research Council (VR) and Knut \& Alice Wallenberg Foundation (KAW), Sweden;
European Organization for Nuclear Research, Switzerland;
Suranaree University of Technology (SUT), National Science and Technology Development Agency (NSTDA) and National Science, Research and Innovation Fund (NSRF via PMU-B B05F650021), Thailand;
Turkish Energy, Nuclear and Mineral Research Agency (TENMAK), Turkey;
National Academy of  Sciences of Ukraine, Ukraine;
Science and Technology Facilities Council (STFC), United Kingdom;
National Science Foundation of the United States of America (NSF) and United States Department of Energy, Office of Nuclear Physics (DOE NP), United States of America.
In addition, individual groups or members have received support from:
Czech Science Foundation (grant no. 23-07499S), Czech Republic;
FORTE project, reg.\ no.\ CZ.02.01.01/00/22\_008/0004632, Czech Republic, co-funded by the European Union, Czech Republic;
European Research Council (grant nos. 101220549, 950692), European Union;
Deutsche Forschungs Gemeinschaft (DFG, German Research Foundation) ``Neutrinos and Dark Matter in Astro- and Particle Physics'' (grant no. SFB 1258), Germany;
CONVECS project, CUP C97H23001700002 FESR 2021-2027 program, Italy.
\end{acknowledgement}

\bibliographystyle{utphys}
\bibliography{paper.bib}

\newpage
\appendix

%
%

\section{The ALICE Collaboration}
\label{app:collab}
\begin{flushleft} 
\small

D.A.H.~Abdallah\,\orcidlink{0000-0003-4768-2718}\,$^{\rm 134}$, 
I.J.~Abualrob\,\orcidlink{0009-0005-3519-5631}\,$^{\rm 112}$, 
S.~Acharya\,\orcidlink{0000-0002-9213-5329}\,$^{\rm 49}$, 
K.~Agarwal\,\orcidlink{0000-0001-5781-3393}\,$^{\rm II,}$$^{\rm 23}$, 
G.~Aglieri Rinella\,\orcidlink{0000-0002-9611-3696}\,$^{\rm 32}$, 
L.~Aglietta\,\orcidlink{0009-0003-0763-6802}\,$^{\rm 24}$, 
N.~Agrawal\,\orcidlink{0000-0003-0348-9836}\,$^{\rm 25}$, 
Z.~Ahammed\,\orcidlink{0000-0001-5241-7412}\,$^{\rm 132}$, 
S.~Ahmad\,\orcidlink{0000-0003-0497-5705}\,$^{\rm 15}$, 
Z.~Akbar$^{\rm 79}$
% ...

\section*{Affiliation Notes}

$^{\rm I}$ Deceased\\
$^{\rm II}$ Also at: INFN Trieste\\
$^{\rm III}$ Also at: Fondazione Bruno Kessler (FBK), Trento, Italy\\
$^{\rm IV}$ Also at: Czech Technical University in Prague (CZ)\\
$^{\rm V}$ Also at: Instituto de Fisica da Universidade de Sao Paulo\\
$^{\rm VI}$ Also at: Dipartimento DET del Politecnico di Torino, Turin, Italy\\
$^{\rm VII}$ Also at: Department of Applied Physics, Aligarh Muslim University, Aligarh, India\\
$^{\rm VIII}$ Also at: Institute of Theoretical Physics, University of Wroclaw, Poland\\
$^{\rm IX}$ Also at: Facultad de Ciencias, Universidad Nacional Aut\'{o}noma de M\'{e}xico, Mexico City, Mexico\\

\section*{Collaboration Institutes}

$^{1}$ A.I. Alikhanyan National Science Laboratory (Yerevan Physics Institute) Foundation, Yerevan, Armenia\\
$^{2}$ AGH University of Krakow, Cracow, Poland\\
$^{3}$ Bogolyubov Institute for Theoretical Physics, National Academy of Sciences of Ukraine, Kyiv, Ukraine\\
$^{4}$ Bose Institute, Department of Physics  and Centre for Astroparticle Physics and Space Science (CAPSS), Kolkata, India\\
$^{5}$ California Polytechnic State University, San Luis Obispo, California, United States\\
$^{6}$ Central China Normal University, Wuhan, China\\
$^{7}$ Centro de Aplicaciones Tecnol\'{o}gicas y Desarrollo Nuclear (CEADEN), Havana, Cuba\\
$^{8}$ Centro de Investigaci\'{o}n y de Estudios Avanzados (CINVESTAV), Mexico City and M\'{e}rida, Mexico\\
$^{9}$ Chicago State University, Chicago, Illinois, United States\\
$^{10}$ China Nuclear Data Center, China Institute of Atomic Energy, Beijing, China\\
$^{11}$ China University of Geosciences, Wuhan, China\\
$^{12}$ Chungbuk National University, Cheongju, Republic of Korea\\
$^{13}$ Comenius University Bratislava, Faculty of Mathematics, Physics and Informatics, Bratislava, Slovak Republic\\
$^{14}$ Creighton University, Omaha, Nebraska, United States\\
$^{15}$ Department of Physics, Aligarh Muslim University, Aligarh, India\\
$^{16}$ Department of Physics, Pusan National University, Pusan, Republic of Korea\\
$^{17}$ Department of Physics, Sejong University, Seoul, Republic of Korea\\
$^{18}$ Department of Physics, University of California, Berkeley, California, United States\\
$^{19}$ Department of Physics, University of Oslo, Oslo, Norway\\
$^{20}$ Department of Physics and Technology, University of Bergen, Bergen, Norway\\
$^{21}$ Dipartimento di Fisica, Universit\`{a} di Pavia, Pavia, Italy\\
$^{22}$ Dipartimento di Fisica dell'Universit\`{a} and Sezione INFN, Cagliari, Italy\\
$^{23}$ Dipartimento di Fisica dell'Universit\`{a} and Sezione INFN, Trieste, Italy\\
$^{24}$ Dipartimento di Fisica dell'Universit\`{a} and Sezione INFN, Turin, Italy\\
$^{25}$ Dipartimento di Fisica e Astronomia dell'Universit\`{a} and Sezione INFN, Bologna, Italy\\
$^{26}$ Dipartimento di Fisica e Astronomia dell'Universit\`{a} and Sezione INFN, Catania, Italy\\
$^{27}$ Dipartimento di Fisica e Astronomia dell'Universit\`{a} and Sezione INFN, Padova, Italy\\
$^{28}$ Dipartimento di Fisica `E.R.~Caianiello' dell'Universit\`{a} and Gruppo Collegato INFN, Salerno, Italy\\
$^{29}$ Dipartimento DISAT del Politecnico and Sezione INFN, Turin, Italy\\
$^{30}$ Dipartimento di Scienze MIFT, Universit\`{a} di Messina, Messina, Italy\\
$^{31}$ Dipartimento Interateneo di Fisica `M.~Merlin' and Sezione INFN, Bari, Italy\\
$^{32}$ European Organization for Nuclear Research (CERN), Geneva, Switzerland\\
$^{33}$ Faculty of Electrical Engineering, Mechanical Engineering and Naval Architecture, University of Split, Split, Croatia\\
$^{34}$ Faculty of Nuclear Sciences and Physical Engineering, Czech Technical University in Prague, Prague, Czech Republic\\
$^{35}$ Faculty of Physics, Sofia University, Sofia, Bulgaria\\
$^{36}$ Faculty of Science, P.J.~\v{S}af\'{a}rik University, Ko\v{s}ice, Slovak Republic\\
$^{37}$ Faculty of Technology, Environmental and Social Sciences, Bergen, Norway\\
$^{38}$ Frankfurt Institute for Advanced Studies, Johann Wolfgang Goethe-Universit\"{a}t Frankfurt, Frankfurt, Germany\\
$^{39}$ Fudan University, Shanghai, China\\
$^{40}$ Gauhati University, Department of Physics, Guwahati, India\\
$^{41}$ Helmholtz-Institut f\"{u}r Strahlen- und Kernphysik, Rheinische Friedrich-Wilhelms-Universit\"{a}t Bonn, Bonn, Germany\\
$^{42}$ Helsinki Institute of Physics (HIP), Helsinki, Finland\\
$^{43}$ High Energy Physics Group,  Universidad Aut\'{o}noma de Puebla, Puebla, Mexico\\
$^{44}$ Horia Hulubei National Institute of Physics and Nuclear Engineering, Bucharest, Romania\\
$^{45}$ HUN-REN Wigner Research Centre for Physics, Budapest, Hungary\\
$^{46}$ Indian Institute of Technology Bombay (IIT), Mumbai, India\\
$^{47}$ Indian Institute of Technology Indore, Indore, India\\
$^{48}$ INFN, Laboratori Nazionali di Frascati, Frascati, Italy\\
$^{49}$ INFN, Sezione di Bari, Bari, Italy\\
$^{50}$ INFN, Sezione di Bologna, Bologna, Italy\\
$^{51}$ INFN, Sezione di Cagliari, Cagliari, Italy\\
$^{52}$ INFN, Sezione di Catania, Catania, Italy\\
$^{53}$ INFN, Sezione di Padova, Padova, Italy\\
$^{54}$ INFN, Sezione di Pavia, Pavia, Italy\\
$^{55}$ INFN, Sezione di Torino, Turin, Italy\\
$^{56}$ INFN, Sezione di Trieste, Trieste, Italy\\
$^{57}$ Inha University, Incheon, Republic of Korea\\
$^{58}$ Institute for Gravitational and Subatomic Physics (GRASP), Utrecht University/Nikhef, Utrecht, Netherlands\\
$^{59}$ Institute of Experimental Physics, Slovak Academy of Sciences, Ko\v{s}ice, Slovak Republic\\
$^{60}$ Institute of Physics, Homi Bhabha National Institute, Bhubaneswar, India\\
$^{61}$ Institute of Physics of the Czech Academy of Sciences, Prague, Czech Republic\\
$^{62}$ Institute of Space Science (ISS), Bucharest, Romania\\
$^{63}$ Institut f\"{u}r Kernphysik, Johann Wolfgang Goethe-Universit\"{a}t Frankfurt, Frankfurt, Germany\\
$^{64}$ Instituto de Ciencias Nucleares, Universidad Nacional Aut\'{o}noma de M\'{e}xico, Mexico City, Mexico\\
$^{65}$ Instituto de F\'{i}sica, Universidade Federal do Rio Grande do Sul (UFRGS), Porto Alegre, Brazil\\
$^{66}$ Instituto de F\'{\i}sica, Universidad Nacional Aut\'{o}noma de M\'{e}xico, Mexico City, Mexico\\
$^{67}$ iThemba LABS, National Research Foundation, Somerset West, South Africa\\
$^{68}$ Jeonbuk National University, Jeonju, Republic of Korea\\
$^{69}$ Korea Institute of Science and Technology Information, Daejeon, Republic of Korea\\
$^{70}$ Laboratoire de Physique Subatomique et de Cosmologie, Universit\'{e} Grenoble-Alpes, CNRS-IN2P3, Grenoble, France\\
$^{71}$ Lawrence Berkeley National Laboratory, Berkeley, California, United States\\
$^{72}$ Lund University Department of Physics, Division of Particle Physics, Lund, Sweden\\
$^{73}$ Marietta Blau Institute, Vienna, Austria\\
$^{74}$ Nagasaki Institute of Applied Science, Nagasaki, Japan\\
$^{75}$ Nara Women{'}s University (NWU), Nara, Japan\\
$^{76}$ National Centre for Nuclear Research, Warsaw, Poland\\
$^{77}$ National Institute of Science Education and Research, Homi Bhabha National Institute, Jatni, India\\
$^{78}$ National Nuclear Research Center, Baku, Azerbaijan\\
$^{79}$ National Research and Innovation Agency - BRIN, Jakarta, Indonesia\\
$^{80}$ Niels Bohr Institute, University of Copenhagen, Copenhagen, Denmark\\
$^{81}$ Nikhef, National institute for subatomic physics, Amsterdam, Netherlands\\
$^{82}$ Nuclear Physics Group, STFC Daresbury Laboratory, Daresbury, United Kingdom\\
$^{83}$ Nuclear Physics Institute of the Czech Academy of Sciences, Husinec-\v{R}e\v{z}, Czech Republic\\
$^{84}$ Oak Ridge National Laboratory, Oak Ridge, Tennessee, United States\\
$^{85}$ Ohio State University, Columbus, Ohio, United States\\
$^{86}$ Physics department, Faculty of science, University of Zagreb, Zagreb, Croatia\\
$^{87}$ Physics Department, Panjab University, Chandigarh, India\\
$^{88}$ Physics Department, University of Jammu, Jammu, India\\
$^{89}$ Physics Program and International Institute for Sustainability with Knotted Chiral Meta Matter (WPI-SKCM$^{2}$), Hiroshima University, Hiroshima, Japan\\
$^{90}$ Physikalisches Institut, Eberhard-Karls-Universit\"{a}t T\"{u}bingen, T\"{u}bingen, Germany\\
$^{91}$ Physikalisches Institut, Ruprecht-Karls-Universit\"{a}t Heidelberg, Heidelberg, Germany\\
$^{92}$ Physik Department, Technische Universit\"{a}t M\"{u}nchen, Munich, Germany\\
$^{93}$ Politecnico di Bari and Sezione INFN, Bari, Italy\\
$^{94}$ Research Division and ExtreMe Matter Institute EMMI, GSI Helmholtzzentrum f\"ur Schwerionenforschung GmbH, Darmstadt, Germany\\
$^{95}$ Saga University, Saga, Japan\\
$^{96}$ Saha Institute of Nuclear Physics, Homi Bhabha National Institute, Kolkata, India\\
$^{97}$ School of Physics and Astronomy, University of Birmingham, Birmingham, United Kingdom\\
$^{98}$ Secci\'{o}n F\'{\i}sica, Departamento de Ciencias, Pontificia Universidad Cat\'{o}lica del Per\'{u}, Lima, Peru\\
$^{99}$ SUBATECH, IMT Atlantique, Nantes Universit\'{e}, CNRS-IN2P3, Nantes, France\\
$^{100}$ Sungkyunkwan University, Suwon City, Republic of Korea\\
$^{101}$ Suranaree University of Technology, Nakhon Ratchasima, Thailand\\
$^{102}$ Technical University of Ko\v{s}ice, Ko\v{s}ice, Slovak Republic\\
$^{103}$ The Henryk Niewodniczanski Institute of Nuclear Physics, Polish Academy of Sciences, Cracow, Poland\\
$^{104}$ The University of Texas at Austin, Austin, Texas, United States\\
$^{105}$ Universidad Aut\'{o}noma de Sinaloa, Culiac\'{a}n, Mexico\\
$^{106}$ Universidade de S\~{a}o Paulo (USP), S\~{a}o Paulo, Brazil\\
$^{107}$ Universidade Estadual de Campinas (UNICAMP), Campinas, Brazil\\
$^{108}$ Universidade Federal do ABC, Santo Andre, Brazil\\
$^{109}$ Universitatea Nationala de Stiinta si Tehnologie Politehnica Bucuresti, Bucharest, Romania\\
$^{110}$ University of Cape Town, Cape Town, South Africa\\
$^{111}$ University of Derby, Derby, United Kingdom\\
$^{112}$ University of Houston, Houston, Texas, United States\\
$^{113}$ University of Jyv\"{a}skyl\"{a}, Jyv\"{a}skyl\"{a}, Finland\\
$^{114}$ University of Kansas, Lawrence, Kansas, United States\\
$^{115}$ University of Liverpool, Liverpool, United Kingdom\\
$^{116}$ University of Science and Technology of China, Hefei, China\\
$^{117}$ University of Silesia in Katowice, Katowice, Poland\\
$^{118}$ University of South-Eastern Norway, Kongsberg, Norway\\
$^{119}$ University of Tennessee, Knoxville, Tennessee, United States\\
$^{120}$ University of the Witwatersrand, Johannesburg, South Africa\\
$^{121}$ University of Tokyo, Tokyo, Japan\\
$^{122}$ University of Tsukuba, Tsukuba, Japan\\
$^{123}$ Universit\"{a}t M\"{u}nster, Institut f\"{u}r Kernphysik, M\"{u}nster, Germany\\
$^{124}$ Universit\'{e} Clermont Auvergne, CNRS/IN2P3, LPC, Clermont-Ferrand, France\\
$^{125}$ Universit\'{e} de Lyon, CNRS/IN2P3, Institut de Physique des 2 Infinis de Lyon, Lyon, France\\
$^{126}$ Universit\'{e} de Strasbourg, CNRS, IPHC UMR 7178, F-67000 Strasbourg, France, Strasbourg, France\\
$^{127}$ Universit\'{e} Paris-Saclay, Centre d'Etudes de Saclay (CEA), IRFU, D\'{e}partment de Physique Nucl\'{e}aire (DPhN), Saclay, France\\
$^{128}$ Universit\'{e}  Paris-Saclay, CNRS/IN2P3, IJCLab, Orsay, France\\
$^{129}$ Universit\`{a} degli Studi di Foggia, Foggia, Italy\\
$^{130}$ Universit\`{a} del Piemonte Orientale, Vercelli, Italy\\
$^{131}$ Universit\`{a} di Brescia, Brescia, Italy\\
$^{132}$ Variable Energy Cyclotron Centre, Homi Bhabha National Institute, Kolkata, India\\
$^{133}$ Warsaw University of Technology, Warsaw, Poland\\
$^{134}$ Wayne State University, Detroit, Michigan, United States\\
$^{135}$ Yale University, New Haven, Connecticut, United States\\
$^{136}$ Yildiz Technical University, Istanbul, Turkey\\
$^{137}$ Yonsei University, Seoul, Republic of Korea\\
$^{138}$ Affiliated with an institute formerly covered by a cooperation agreement with CERN\\
$^{139}$ Affiliated with an international laboratory covered by a cooperation agreement with CERN.\\

\end{flushleft} 

\end{document}